# Barrier Free Internet Access: Evaluating the Cyber Security Risk Posed by the Adoption of Bring Your Own Devices to e-Learning Network Infrastructure

E. T. Tchao
Dept. of Computer Engineering
KNUST

Richard Y. Ansah
University I.T. Services
KNUST

Seth D. Kotey
Dept. of Computer Engineering
KNUST

## ABSTRACT

The adoption of Bring Your Own Device (BYOD) – also known as Bring Your Own Technology (BYOT), Bring Your Own Phone (BYOP), or Bring Your Own Personal Computer (BYOPC) – is a policy which allows people access to privileged resources, information and services available on the private computer network of an organization using their own personal computer devices. BYOD, since its emergence in 2009, courtesy of Intel, is now a common practice in many organizations. Academic institutions that attempt to implement BYOD, can derive many benefits as well as many risks to its network infrastructure, largely security-based.

This paper presents an assessment of a WLAN network which has been deployed for a campus-wide data centric e-learning platform at Kwame Nkrumah University of Science and Technology (KNUST) towards the overall objective of achieving a barrier free internet access to enhance the teaching and learning process at the university. The paper subsequently evaluates the WLAN infrastructure, its accompanying BYOD set-up, and associated likely security risks and threats, and recommends appropriate solutions.

## 1. INTRODUCTION

Computer networks have grown exponentially in popularity, demand, and convenience since the advent of the US DoD's ARPANET in 1969[1]. Today, computer networks afford us the ability to conduct business, share information, stay in touch with professional and personal connections, and access news and entertainment all over the world from the comfort of one's home, classroom, office, or anywhere with an Internet connection [2]. With the arrival of next generation computer systems and networking technology, real time interaction with anyone can be experienced as well as obtaining any information.

The explosion of computer networks, along with many users desiring to stay connected to each other and the world, meant the growth of digital information, which is much more difficult to protect than hard copy files and folders. This makes cyber security inconvenient because there always has to be a compromise between robustness and simplicity. That is, the more robust and secure the security mechanisms, the more inconvenient the process becomes [3]. Moreover, the current trend is to share information, not protect it. Every now and then, news of compromised individuals, schools, research organizations, even governments, from social media to secure website portals is encountered. Still, people will share their data and information on social media, visit questionable websites, and download files from the Internet that probably contain malware.

Over the years since computer networks have no longer became restricted to governments and educational institutions, there have appeared many forms of security issues, and many more ways of ensuring security. There are now various types, forms, versions and mutations of malware, and just as many antimalware, firewalls, authentication protocols, encryption techniques, and so on [4].

The number of public Ethernet and Wi-Fi LANs has increased in the last decade, particularly in schools, university campuses, and offices that want to streamline educational or work activities that require networking by providing their own private networks. Wireless LAN is the choice network in many organizations, public venues, and homes; they are even allowed on aircrafts these days; and they offer a wide range of use cases, deployment scenarios, and security budgets according to the requirements of the network [5]. Usually, the schools and organizations owned all the computers and resources on the network. This made managing and securing the private network more streamlined and effective.

However, in recent years, mobile devices have become so ubiquitous that people no longer bat an eye when an individual operates two or more mobile devices [6]. Modern mobile devices are now equipped with faster microprocessors, better and more memory, and better integration and support in connecting to enterprise services, making them more capable to students and workers to perform their regular functions [7]. There has been an exponential growth of smartphones sales, which according to Gartner's report has led to the reduction of laptop sales. Gartner even goes further to predict that the use of smartphones and tablet computers in educational institutions will replace that of laptops in the very near future [8].

This development has led many educational institutions and organizations to adopt the Bring Your Own Device (BYOD) initiative. Pioneered by Intel in 2009 [9], BYOD means exactly as the name suggests: People can now access private networks of their respective organizations with their own personal devices. This strategy has many benefits to both the organization and the individuals:

1. **Education** – The larger number of mobile device users exist in schools and universities. It is not uncommon to find Wi-Fi in universities these days, accessed by the thousands by students to access course materials, share files, and browse the Internet. [10][11][12][13]
2. **Enterprise** – In a business-oriented environment, connectivity to information is crucial to

productivity. The ability to access and update data from anywhere ensures higher efficiency, convenience, and worker satisfaction.

3. **Healthcare** – In hospitals where many rules and regulations exist, it is the last place one would expect a BYOD network. But many doctors can appreciate this initiative as a convenient way to access hospital data, monitor patients, etc. all from the comfort of their offices and homes.

Also, BYOD offers added benefits of saving an organization equipment costs, offers stronger protection of personal, sensitive information, and mobility [14].

However, BYOD is not without security risks. This is mainly because the clients choose their own devices. In view of this, network architects and administrators often have to make tough choices throughout the process of designing, installing, managing and securing their networks. Among the many factors they have to consider about the network infrastructure are the hardware choices that can adequately accommodate bandwidth requirements, the number of potential clients, types of mobile devices, and the purpose and intentions of the users.

Some benefits and challenges of incorporating a BYOD set-up into an organization's private network have been stated. This case study seeks to assess the cybersecurity risks posed by BYOD initiative currently started by KNUST to afford the students barrier free access to network resources. The study evaluates the implemented network architecture at KNUST and subsequently presents the vulnerability assessment results.

# 2. KNUST NETWORK

## 2.1 KNUST Network Infrastructure Design

Before the security details of KNUST's BYOD set-up are investigated, the existing WAN needs to be looked at, or in this case, the Campus Area Network (CAN) as shown in Figure 1. The entire KNUST wired network backbone rides on a single link to the outside world with a bandwidth of 144 Mbps as of June, 2015. This connectivity is supplied by an ISP, Vodafone Ghana. The link enters the KNUST Network Operations and Infrastructure Department (NOID) via fiber optic cables. The link passes through a cascade of firewalls, caching devices, and core Ethernet switches. Next, the link is split up by distribution switches with more fiber optic and Ethernet cables to the various faculties in KNUST.

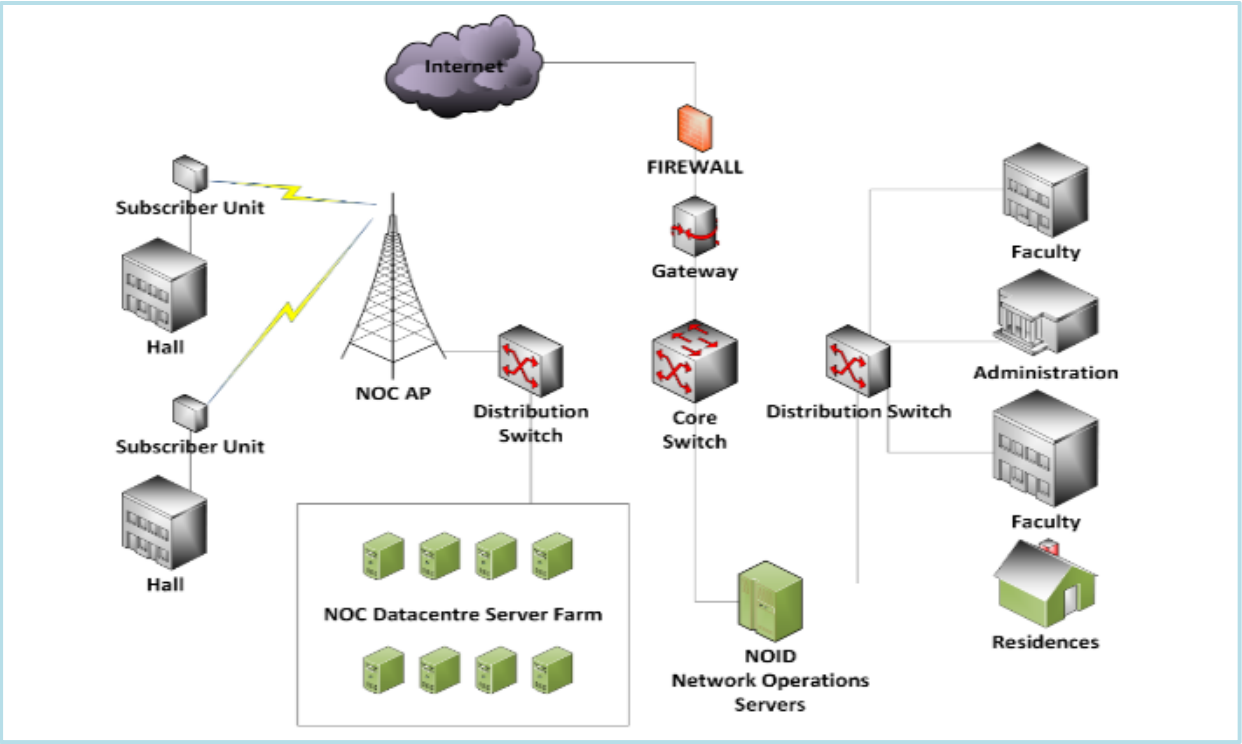

**Figure 1: Simplified Network Diagram of KNUST WAN**

The well-structured Ethernet backbone also support a VoIP system, and are linked to web, DNS, database, directory, and application servers that host KNUST websites, web apps, school management data, and other essential information

## 2.2 KNUST Wireless LAN Infrastructure Design

The robustness of a WLAN depends on a well-designed wired network backbone, which has been covered. The next step was to investigate the requirements of the WLAN used for the BYOD set-up. These drive the basic principles such as coverage, capacity and security.

## 2.3 WLAN Architecture

WLAN architectures are based on the levels (tiers) of the access node. The tiers indicate how many devices sit between the WAN and an access point as shown in Figure 2.

The main types of access points available are Autonomous Access Points (AAP) and Lightweight Access Points (LAP). Autonomous, as the name implies, consists of routers that are autonomous, i.e. they care completely self-sufficient, standalone devices that can connect multiple clients to a central wired LAN network [15]. An LAP, on the other hand, has to be controlled by an AAP or premise-based WLAN controller for scalability, added maintenance, and Quality of Service.

Looking at AP installations around KNUST, it is noticed that both AAPs (which can be seen mounted on the side of the buildings) and LAPs (which reside in the classrooms and lecture halls) are logically connected in a 2-tier configuration, shown in Figure 2. The 2-tier configuration allows for maximum coverage and allows for simpler allocation of the same SSID for many APs.

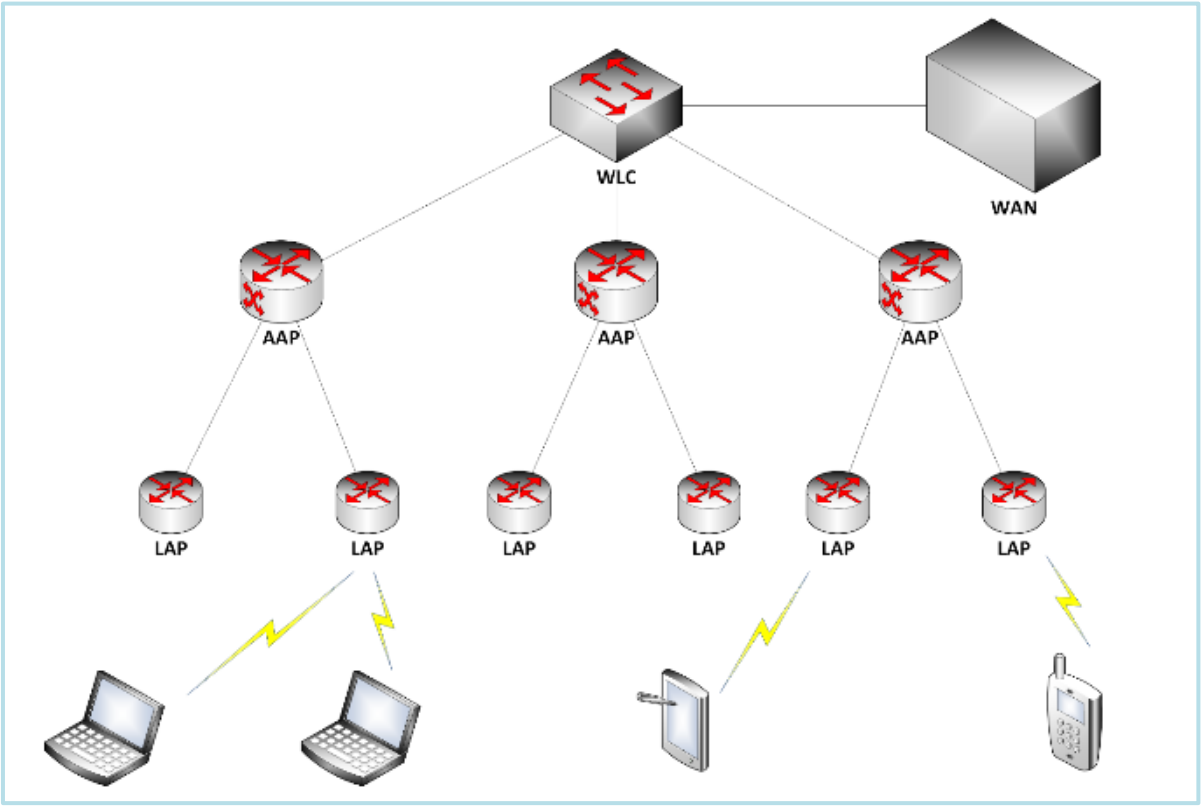

**Figure 2: Logical 2-Tier Access Node Configuration**

## 2.4 WLAN Access Point Device Types

APs come in single channel (rated at 2.4 GHz) and dual channel (rated at 2.4 GHz and 5 GHz). The AP type often depends on which types of IEEE 802.11 technologies it can support. The APs around KNUST are dual channel, each promising a theoretical bandwidth of 300Mbps[1] per channel, and capable of supporting legacy technologies 802.11b and 802.11g as well as the newer 802.11a and 802.11n [16].

Of course, factors like signal power, channel allocation and interference, channel bonding, distance constraints and obstacles contribute to increases or decreases in bandwidth and throughput [7]. Typical throughputs are much less. Considering a dual channel AP with theoretical throughput of 300Mbps per channel would yield 600Mps. Yet, in practice, several factors drastically reduce this figure.

- Protocol and packet overhead can reduce throughput by 40 to 50%
- Slow or "far away" clients sending a packet at 1Mps would take 100 times more time than a client sending the same packet at 100Mps
- Uneven distribution of clients, all contending for access, even with dual channels, can reduce throughput by another 50%.
- Other factors such as packet retransmission, channel interference from rouge networks, etc. contribute to a further 25% reduction in throughput.

Eventually, the actual throughput per client is just about 2 to 3 Mbps on a good day [17].

**Table 1. Basic Wi-Fi Requirements**

| REQUIREMENT CATEGORY | CASES AND CONDITIONS |
|---|---|
| **Users** | 1. Students<br>2. Staff |
| **Device support** | 1. Desktops<br>2. Laptops<br>3. Tablets<br>4. Mobile phones |
| **Wi-Fi Technologies** | 1. IEEE 802.11b/g (2.5 GHz[2])<br>2. IEEE 802.11a/n (5.0 GHz) |
| **Coverage** | 1. Classrooms<br>2. Lecture halls<br>3. Public venues |
| **User Density** | 1. Classrooms: 10 – 500<br>2. Lecture halls: 20 – 1000<br>3. Public venues: 100 – 2000 |
| **Availability Time** | 1. 24 hours<br>2. 7 days |
| **Applications** | 1. Internet<br>2. File sharing<br>3. Media streaming<br>4. Application protocols (HTTP, FTP, P2P, etc.)<br>5. Transport protocols (TCP, UDP) |

## 2.5 KNUST WLAN attributes

The attribute of the deployed WLAN as seen on a laptop computer running Windows 10 Pro is as summarized:

a. *Wireless Properties*

Name/SSID – WIFI
Network Type – Access Point
Network Availability – All Users
Network Security – No Authentication (Open)
Encryption Type – None
IPv4 Connectivity – Internet
IPv6 Connectivity – No Network Access
Speed – 65 Mbps[3]

b. *Wireless Details*

DHCP Enabled – Yes
DNS Suffix – knust.edu.gh
Default Gateway – 10.9.0.5

## 2.6 KNUST WLAN and BYOD Test Runs

### *2.6.1 First Version*

In early 2015, the initiative to introduce a new WLAN into KNUST was kept under wraps, but known only by rumor to Electrical, Telecommunication, and Computer engineering students. Soon enough, workers began digging trenches and laying network and power cables around campus. The only reliable Wi-Fi available in KNUST were Cloud Ghana, whose APs were only available in halls and hostels but not faculties and Vodafone, whose WLAN signal range was limited to the vicinity of their Internet café. The KNUST WLAN, with SSID *WIFI-KNUST,* did not become operational and accessible to students until sometime in November, 2015. The new network had no security features that were visible to clients, no restrictions to the web domains, and no caps on bandwidth had been imposed. Moreover, users experienced throughputs from 512 kbps to over 5.2 Mbps depending on the time of day, number of connected clients, user's Wi-Fi technology type, and AP signal strength. At the time, the greater population of students were unaware of the new network, and only a few faculties had APs installed on their premises.

### *2.6.2 Second Version*

After a month, the network administrator introduced some restrictions on domains and protocols. Students could not access some websites such as YouTube, and domains that hosted websites with pornographic content; any attempt to access such domains would only redirect the browser to the KNUST website. Also, clients could download files from the Internet using HTTP and HTTPS only; no FTP or P2P connections were allowed.

On the bright side, users were experiencing throughputs up to 256 kbps to over 5.2 Mbps depending on the time of day, number of connected users, the users' Wi-Fi technology type, and AP signal strength. Still many students were unaware of WIFI-KNUST.

### *2.6.3 Third Version*

After a while, clients were no longer restricted to access to any website domain they preferred, nor were they limited by protocols.

However, a second logical network, with SSID *KNUST WIFI SEC* was created. This network was indeed an isolation of *WIFI-KNUST,* a test network. This network was closed off behind an undisclosed proxy firewall with no access list, making it possible for the network administrator to test all manner of security policies that could be deployed on the main network.

### *2.6.4 Fourth Version*

The network administrators decided to implement a security system – the gateway proxy (with IP 10.5.0.7 on port 3905) running an open source captive portal, CoovaChilli [18], to deny unauthorized Internet access, and to redirect browsers to a login portal, similar to the login website of KNUST Student Portal (KNUST, 2016). Internet access was disallowed until a student verified their identity with their school issued student ID and password.

The portal was inefficient because the gateway was not always reliable to:

- Redirect all URLs to the CoovaChilli captive portal
- Authenticate user credentials (host unavailability and timeouts were frequent)
- Access time ended every 10 minutes or so, and the student had to log in again

The addition of the captive portal, in addition to the reduced throughput even at night when there were fewer users (perhaps by a bandwidth cap imposed on the network by the administrators) caused much frustration and disappointment among students with the Wi-Fi service.

### *2.6.5 Current Version*

The captive portal was discontinued. *WIFI-KNUST* is now open, without any authentication. However, *KNUST WIFI SEC,* while also without any WEP, WPA2, or 802.1X encryption or authentication, is fully monitored, and has a better, but not fully functional, captive portal.

## 2.7 KNUST BYOD Security Risks

**1. No Wi-Fi Encryption or Authentication**

Generic WLAN networks implement security authentication types such as:

(i) **WEP** – Encryption for IEEE 802.11 Wi-Fi networks to provide similar privacy of an Ethernet LAN. WEP is very insecure and contains serious security flaws; it is not intended to be the only security measure; superseded by WPA [19][20][21]

(ii) **WPA** – Authentication type for Wi-Fi. Superseded by WPA2 [22].

(iii) **WPA2** – Improved version of WPA (known as WPA2 Personal or WPA2 Enterprise). It is based on a modified form of AES encryption, but still very vulnerable [22][23]

None of these IEEE 802.11 security protocols should be relied on exclusively to secure a WLAN. Still, *WIFI-KNUST* does not implement any of the above.

An open Wi-Fi with no encryption or authentication allows outsiders to use resources such as an Internet connection. It is susceptible to network privacy intrusions like eavesdropping.

**2. Automatic Host Configuration (DHCP)**

Public networks with DHCP disabled are tedious to use and manage since users have to manually choose their own IP addresses, and know extra information like IP addresses of DNS servers and gateway routers. Users may choose incorrect network and DNS Server addresses or conflicting host identifiers. Nevertheless, making such information private provides a sort of security by "IP anonymity." One cannot use a network without correct configuration settings.

Of course, this is not practical for a network for thousands of users in KNUST. So, KNUST's WLANs have DHCP enabled. Therefore, any host may connect to *WIFI-KNUST* and, with no further settings, obtain direct access to the Internet.

**3. Captive Portal**

Although it was an imperfect application, the reasons that led to the recent discontinuation of the captive portal are not fully known, but at least it offered some degree of air gap between *WIFI-KNUST* users and the Internet. Without the portal, there is not much standing between a client and a malicious attacker on the same WLAN or from the Internet.

**4. Device Discovery**

By default, most computers trust Ethernet (mark as private) and distrust Wi-Fi (mark as public). Nevertheless, it is quite easy to change this setting, especially on a Windows operating system. The ability for PCs to discover other PCs on a network is a vital precondition for many network attacks and spreading of malware. Limiting interactions between hosts on a network can help protect them especially keep malware from running rampant [24].

# 3. METHODOLOGY

There are many highly reputable organizations that design and build ultramodern computer networks for individuals and other organizations. Their solutions make them highly acclaimed the world over, and most do not reveal their "trade secrets." They, however, do publish books, whitepapers and reports on many tested and verified practices in computer networks.

In this section, a few of these publications concerning WLAN and BYOD from illustrious conglomerates like Cisco, Aruba, Fortinet, Global Sign, and Intel, are briefly examined and how to use their advice to solve KNUST BYOD challenges.

**1. Trust Your Users – Intel**

As a company that manufactures several billions of microprocessors yearly, Intel does not just allow people to use their own devices. It welcomes and encourages it. So, while many network architects and IT managers worry about user intent, Intel does something radical – it trusts its employees to patronize and adhere to its BYOD program.

This trust is not absolute, of course. This is not a passive dismissal of reality of threats posed by malicious users, but an awareness of it. This translates to:

1. Encouraging BYOD! Intel asserts that clients demeaning a BYOD program, and refusing to use, a Wi-Fi network is just as unfortunate as having security issues. So, as much as possible, KNUST BYOD program should be attractive and exciting.
2. Making the BYOD program as convenient as possible while implementing robust security and privacy measures. An example is making users control the implementation process by being able to choose the access and security level they require to use the corporate network.
3. Regulating what users can and cannot have access to. Users often feel that IT managers highly overestimate the access they have. Besides, the larger percentage of users always use company Wi-Fi for work, and never out of desire to steal information or wreak havoc.

4. Creating separate spaces for personal and company data. This obviously refers to separating storage spaces, databases, and perhaps using different servers to host them [25].

Intel believes that its entire BYOD policy would collapse without user's interest and trust; the belief that security must start with the employee [26].

**2. Enterprise Mobility – Cisco**

Cisco Systems Inc. is a world leader in enterprise network Solutions Corporation. The flourishing of BYOD campaigns was not lost on them. Their Enterprise Mobility solution offers true device freedom without compromising the corporate network [7]. Their solutions offer an organization control of the network when it does not control the devices that use it [27]. Cisco achieves this for the networks its experts design by the following guidelines:

1. Secure Wireless Network Infrastructure – The foundation of a successful BYOD solution is providing an excellent user experience while minimising risk. As iterated over again, this involves building atop a robust Ethernet backbone and using high quality WLCs and APs.
2. Automated Enforcement of Access Policies – This involves automation of authorisation and authentication processes of the devices used on the network, the clients who own them, and the services they wish to access.
3. Web Security – Most computer threats spread when users visit websites. A BYOD solution such as KNUST's should consider utilities that deal with URL filtering, malicious code detection and filtering, and application controls for web-based applications.
4. Low Management Overhead – Managing a campus Wi-Fi and BYOD solution can be made more secure, efficient, and less demanding if there is a centralised management station.
5. Mobile Device and Mobile Application Management – Many enterprises have adopted MDM/MAM – an all-inclusive management of various device types, platforms, applications, user roles and locations, etc. MDM/MAM seeks to make do on the promise of diversity of devices and mobility of users in a BYOD implementation. [27]

To this end, Cisco developed various solutions such as the Cisco Identity Services Engine (ISE – a powerful but flexible and fully customizable platform that addresses all these requirements); Cisco AnyConnect® Secure Mobility Client (a VPN access solution); Cisco ASA CX Context-Aware Firewall and S-Series Web Security Appliance.

**3. Focus on Productivity – Aruba Networks**

Many organizations prefer the perspective the BYOD is a means of increasing productivity to the convenience it offers users and IT managers when managing a WLAN infrastructure. In spite of the obvious benefits of BYOD, Aruba recommends a thorough assessment of your organizational needs, resources to determine if a BYOD solution will benefit the organization. If the assessment predicts impediment to productivity, it is advisable to abandon the initiative [28].

**4. Security is Critical – EYGM**

This is a point worth overemphasis yet. The whole idea of clients bringing their own devices originates from a desire to protect themselves from external scrutiny. People are more likely to check emails, share files, create documents, and work harder using their own devices than using devices supplied – and no doubt controlled – by a company. Hence, while ensuring convenience, it is also important to consider various factors such as device profile, organizational risks, security solutions, deployment scenarios, future state scenarios, mobile device management, etc. As much as possible, create an airtight BYOD system [29].

# 4. KNUST BYOD RECOMMENDATIONS

*1 – Ethernet LAN Performance and Reliability*

First, ensure a robust Ethernet backbone, complete with modern, quality network hardware (high bandwidth routers and switches, Category 5e or better Ethernet and fiber optic cables, UTMs, next generation, context-aware firewalls, etc.). An excellent wired LAN with better traffic and security management will ensure a decent WLAN implementation [7]. The relationship attributes are summarized in Figure 3.

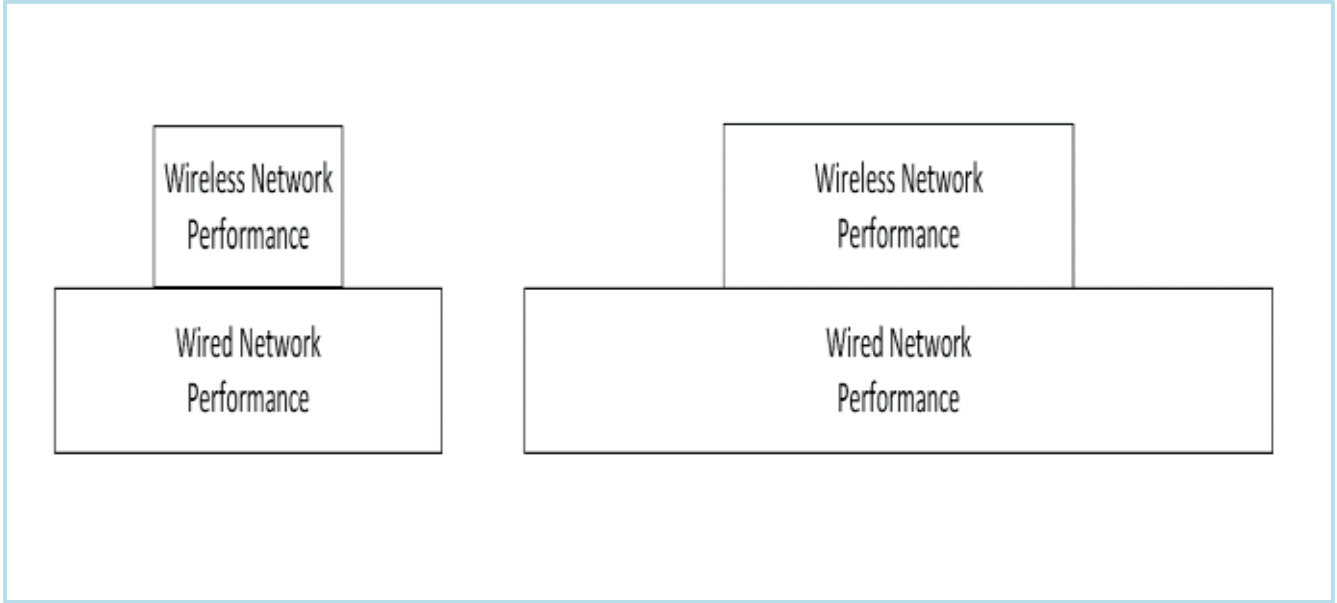

**Figure 3: Relationship between Wired and Wireless Networks**

*2 – WLAN Performance and Reliability*

Secondly, perform requirement assessments and surveys to determine WLAN constraints, as have been discussed in section 3.

Leverage the stability of the underlying wired LAN to build a reliable WLAN [7] using a suitable hierarchical topology for better management and scalability [26].

Use high load dual channel routers and APs that can use channel bonding and multiple data rates, and use PoE so they can be easily placed at vantage points on premises [15]. A fewer number of AAPs is recommended whiles providing LAPs for more access locations as are needed, and choose strategic locations for the APs to ensure maximum coverage and signal power [30], and reduce interferences and obstacles [17][31]. Use the same SSID for multiple APs in a zone to make roaming easy [15].

*3 – Consider Better Network Protocols*

IPv4 has been around for a while – has been the default network addressing protocol for many years.

But that is changing, with IPv4 addresses exhausted and more organizations are switching to IPv6 [1][32]. The performance of a WLAN can be greatly increased by configuring IPv6 network access (alongside IPv4 for devices incapable of using IPv6). IPv4 and IPv6 can be deployed in a number of ways for both independence and interoperability, the easiest and most common being the IPv4v6 Dual Stack Model.

In the End-To-End Dual Stack model, shown in Figure 4, both IPv4 and IPv6 can work fully without interfering with each other if both the access block (end users) and datacenter block (administration end) are both fully capable of deploying IPv6; otherwise, the Hybrid Model with ISATAP and manual tunneling can be used if either the end users or the wired backbone cannot leverage IPv6 [33].

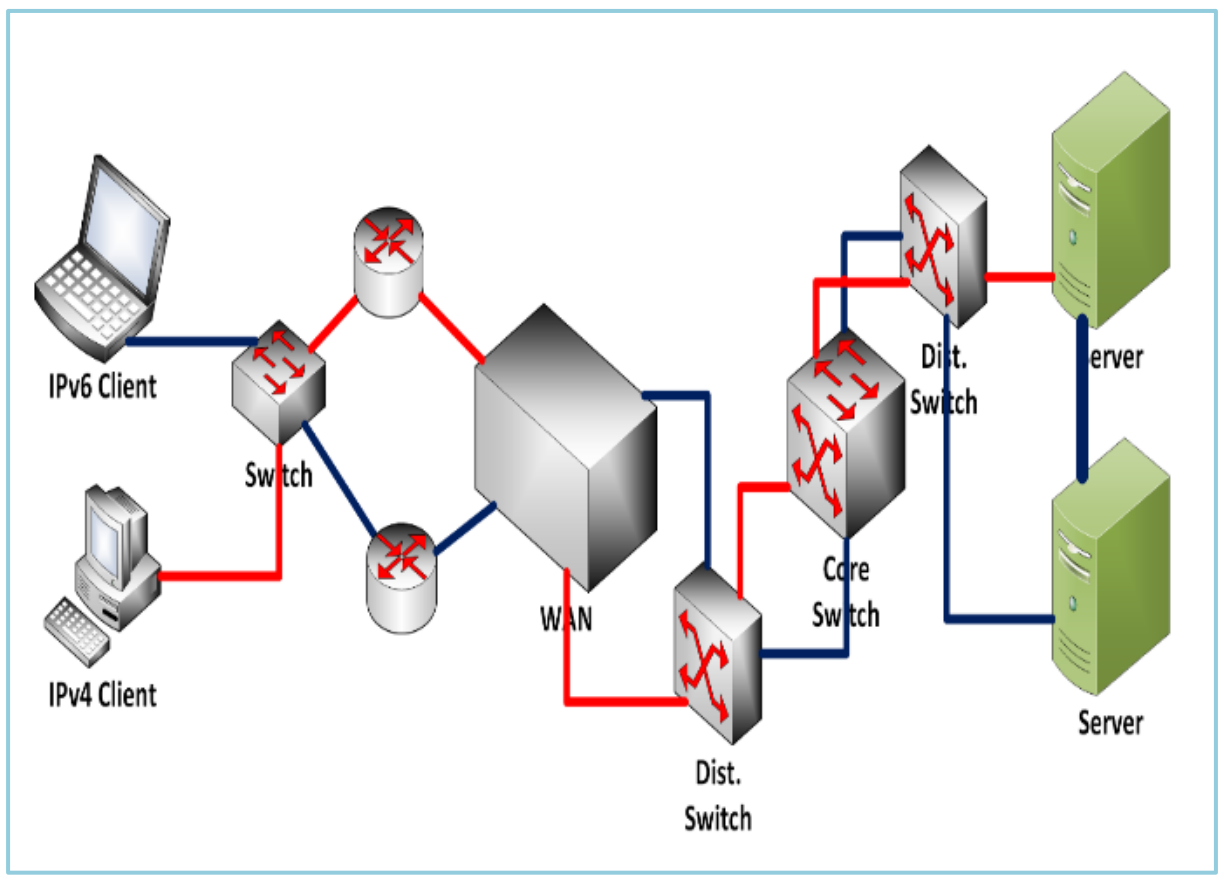

**Figure 4: IPv4v6 End-to-End Dual Stack Model**

IPv6 has many performance and security enhancements over IPv4.

1. IPv6 was designed for faster and more secure host configuration with DHCPv6, or Auto-configuration using a device's MAC address. Experts say this makes IPv6 addresses impossible to spoof [32].
2. IPv6 was also designed for better routing. Improved IP headers prevent IP fragmentation attacks [1]; improved packet size makes it unnecessary for routers to fragment or combine IP packets in order to make it compatible to the underlying LLC and MAC sublayers [1][4[32]].

IPv6 is also better for addressing subnetworks and creating DMZs without the NAT overhead in IPv4 as shown in Figure 5.

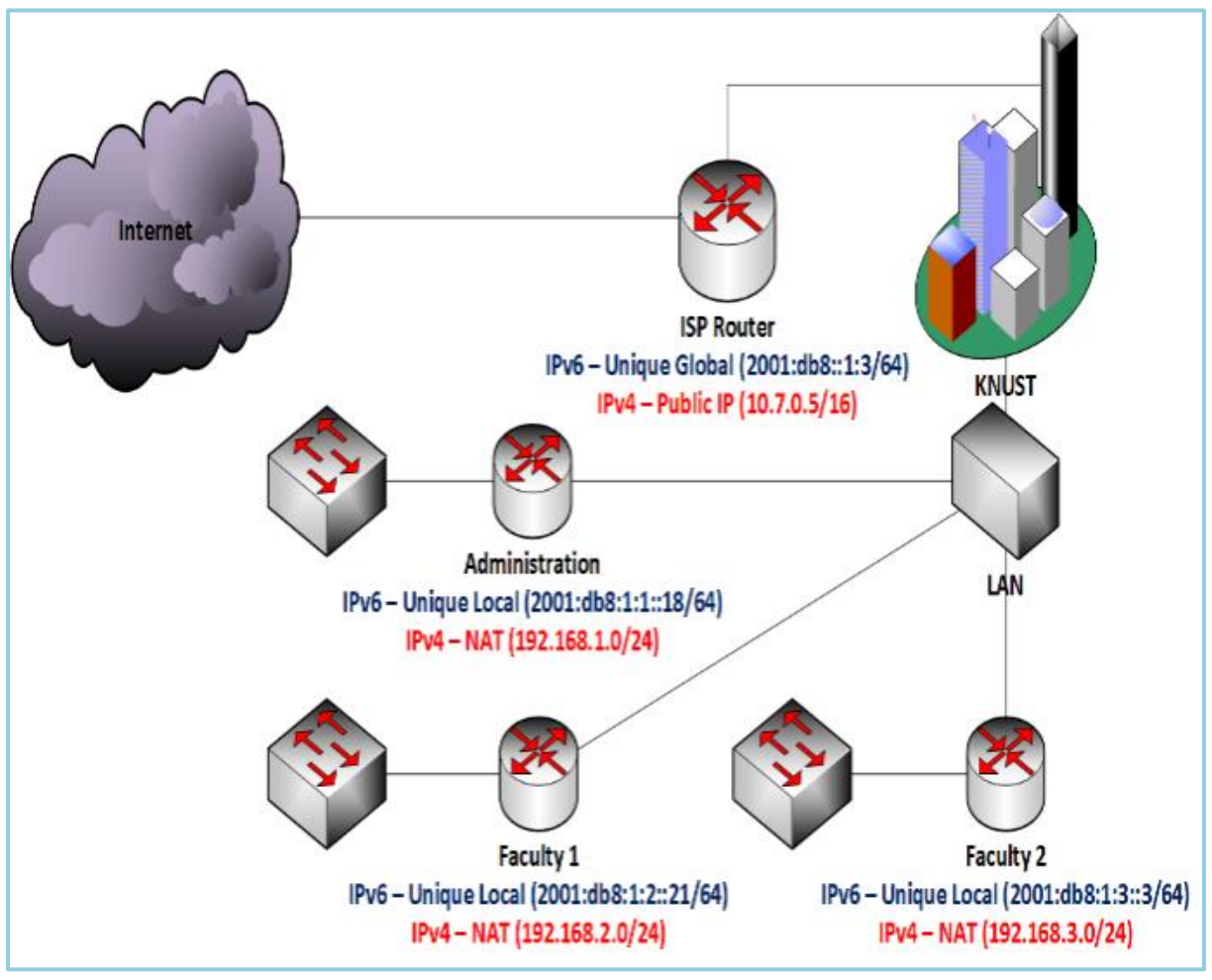

**Figure 5: IPv4 vs IPv6 Global and Subnet Routing**

*4 – Consider Network Segregation*

Network segregation is a good practice, if done correctly, to ensure that various sections of the school's network are isolated and protected from one another and from the Internet [24][4]. Key resources (web, application, and database servers) can be properly isolated and firewalled, and faculties, departments and buildings can be logically separated from one another and from the Internet [34].

In this regard, as already mentioned in the 3rd recommendation, IPv6 is better at dividing corporate networks into subnets, isolations, and DMZs according to the logical structure and needs of KNUST without the extra overhead incurred employing NAT and VLANs in IPv4 [1][32].

*5 – Take Advantage of Basic Wi-Fi Security Options*

The IEEE 802.11 standard Wi-Fi security options discussed in 5 are certainly not ideal, but they can be used in some isolated subnets as additional security layer. WEP and WPA2 are vulnerable, but regular Wi-Fi users hardly ever attempt it [12].

*6 – Monitor and Control the WLAN*

Network administrators should be able to efficiently monitor and manage their networks [35]. Right from the very beginning of the project, integrate a network monitoring system into WLAN management structure. Start with SNMP and work up to sampling various free and retailed reputable network monitoring utilities such as Wireshark, the Dude, Logic Monitor, etc. and sticking with one or more that accomplishes the task. The administrator can also step up to infrastructure monitoring of DHCP and DNS services, etc.

Data collected from these monitoring activities can be analyzed to determine usage of the network infrastructure, bandwidth demand, throughput performance, behavior and activity of users, suspicious behavior, etc. Traffic monitoring can inform many decisions of bandwidth allocation, IP address Management (IPAM), URL restrictions, and so on.

*7 – Implement User Authentication and Mobile Device Management*

Security is critical in a private WLAN, especially one running a BYOD program. It is highly recommended to take a pervasive security measures to ensure that only authorized users can access the right resources: Internet, VoIP, web applications, or storage. Use as many security technologies, protocols, and solutions as reasonably possible. Employing several systems to handle various security scenarios, or a complete security solution such as a UMT.

Design for devices that implement minimum security capabilities, and accept that not all devices will be supported, because considering some legacy devices may very well lead to choices that will compromise the infrastructure. In the meantime, beware of, and take advantage of security capabilities of modern devices (e.g. certificates, dynamic MAC addresses, IPv6 Auto configuration) [36].

Implement MDM system capable of registering, classifying, and managing devices of users through MAC addresses, operating systems, Wi-Fi technology, IP version support, etc.

Lastly, implement captive portals customized for credential-based user authentication, service selection and access level; deploy NAC if there is the need to assess and enforce device security requirements [6].

## 4.1 Proposed Design for KNUST WLAN BYOD

Adopted from Cisco Validated Designs and Cisco Reference Network Architecture [31], a KNUST BYOD solution based on the existing WAN, shown in Figure 6, was designed. The design includes ACMs, Threat Managers, and other network modifications that shall be elaborated now.

*Solution Components*

1. Routers, switches and Access Points that use recent Ethernet and 802.11 technology, e.g.:
   - (i) Cisco Catalyst 2000-X, 3000, and 4000-E Series switches
   - (ii) Cisco vWLC Wireless LAN Controller
2. Access Control Managers (MDM and NAC solutions), e.g.:
   - (i) Cisco Identity Services Engine (ISE)
   - (ii) Cisco Secure Access Control Server (ACS)
   - (iii) Cisco Authorization, Authentication Accounting (AAA) Server
3. Threat Managers, e.g.:
   - (i) Cisco Wireless Intrusion Prevention System (wIPS)
4. BYOD Managers, e.g.:
   - (i) Cisco Mobility Services Engine (MSE)
   - (ii) Cisco Connected Mobile Experiences (CMX)

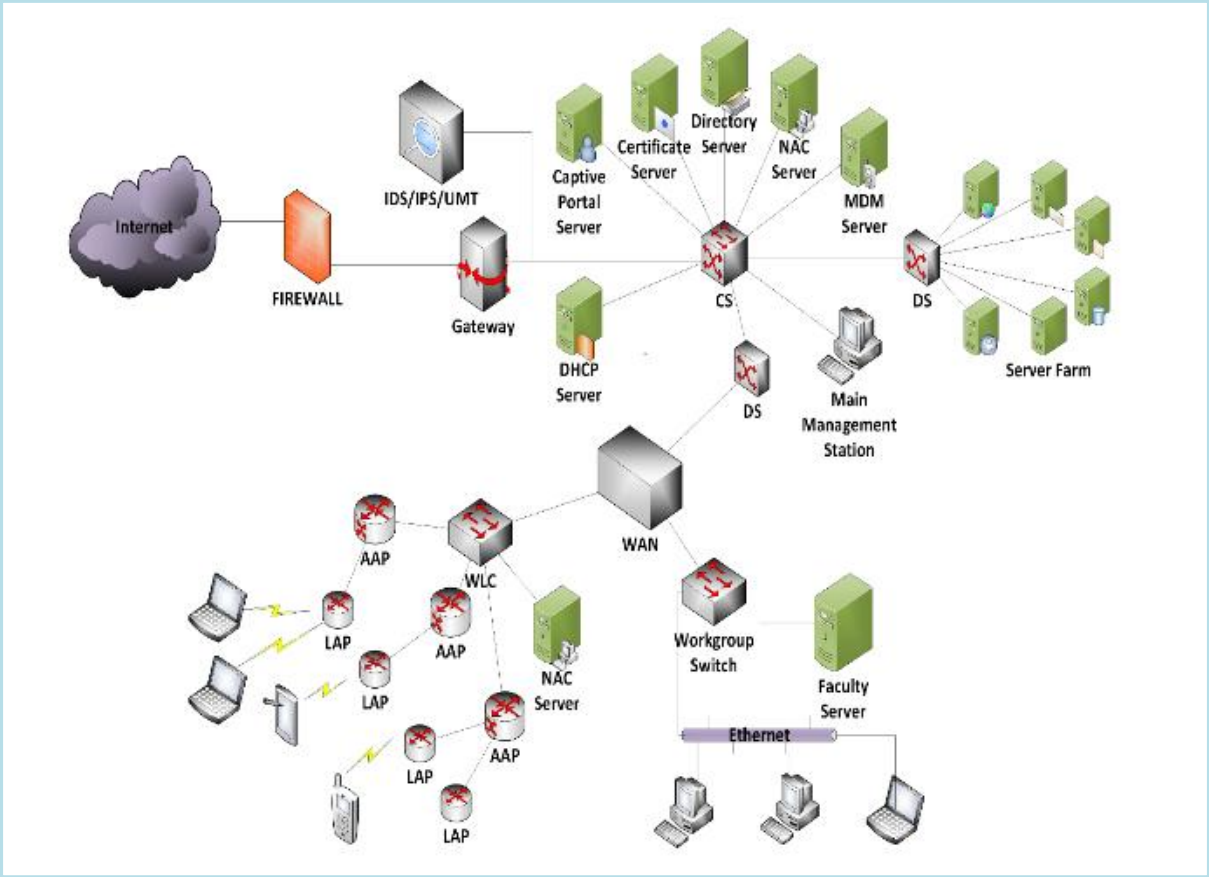

**Figure 6: Design for NUST WAN and BYOD**

*Technology Use Cases*

1. **Subnetting and isolation (DMZs)**
   An educational institution such as KNUST has resources that are intended for authorized users everywhere (Web, DNS, etc.) and resources that should be accessed from the inside the private network only (Email, Databases, VoIP, Captive portals, DHCP, etc.) Such resources can be isolated behind firewalls in DMZs with stringent access control settings. The firewall's ACL allows only IPs belonging to the private network access to certain resources inside the private DMZ, and blocking all others. The public DMZ contains data that when compromised, will not adversely affect the security of KNUST, and be accessed by all as shown in Figure 7.
   The one important property of the DMZs is there are no hosts, dynamic or static, other than the servers located there. All management actions must go through the firewall, ensuring that no backdoors are accidentally built into them.

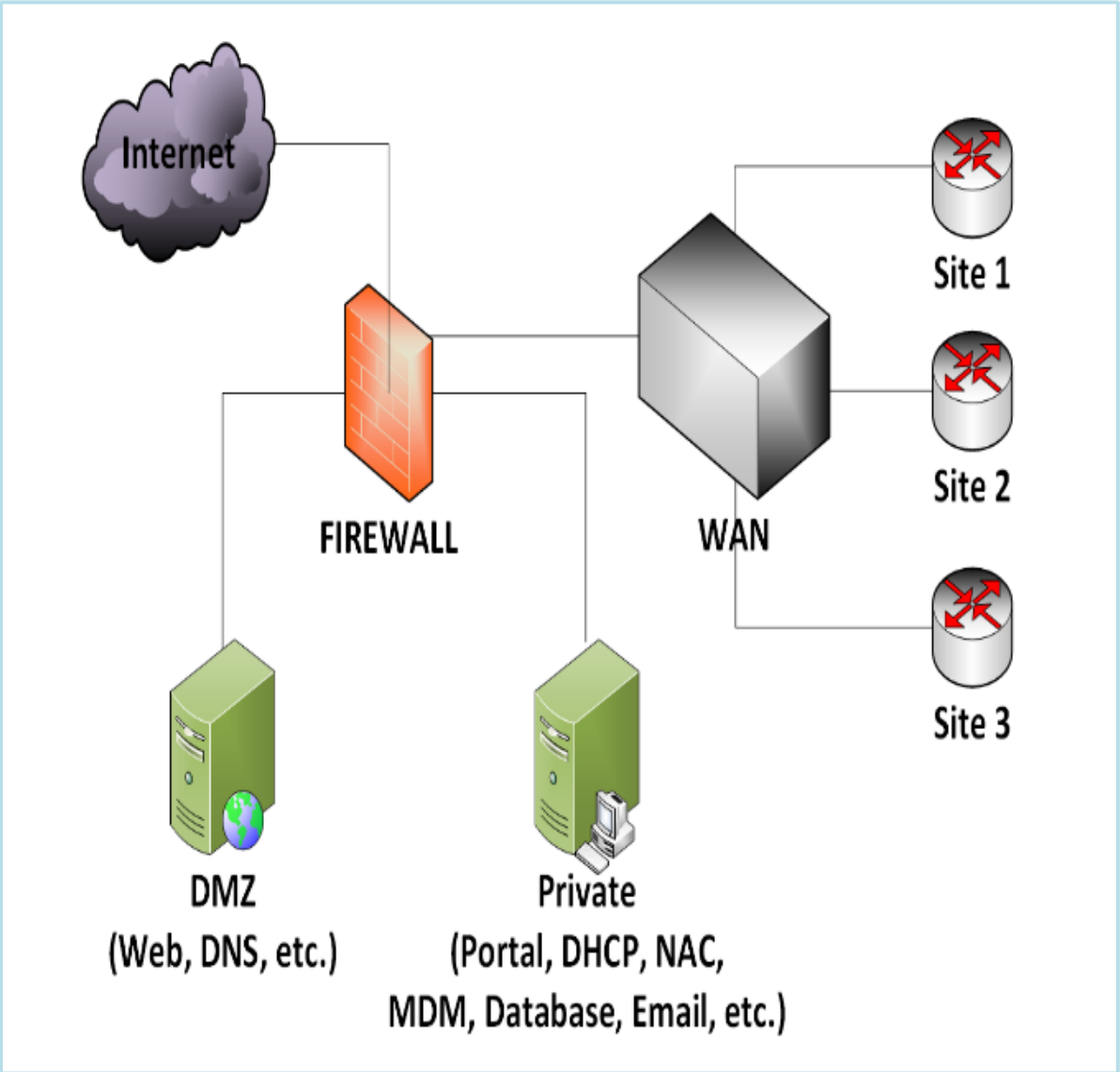

**Figure 7: Firewalled DMZs in a Private Network**

2. **Network Access Control (NAC)**
   NAC is an industrial term to describe the integration and management of several network security control solutions. NAC, as shown in Figure 8, deals with several facets of security such as mobile device and mobile application management (MDM/MAM), antimalware, firewalls and intrusion detection, certificates, credentials and IP address management (IPAM). Often, these systems involve servers that run special applications that communicate with one another and to the NAC server, usually via RPC and ICMP.

3. **Mobile Device Management (MDM)**
   MDM is the administration of networked devices. The NAC system usually controls the MDM's activities. The MDM is responsible for identifying every single computer device connected to the network by collecting data such as:
   - (i) Device name
   - (ii) Device type
   - (iii) Device serial number and/or IMEI
   - (iv) MAC address
   - (v) Device manufacturer
   - (vi) Date manufactured
   - (vii) OS or firmware version and date
   - (viii) IP version (and IP addresses used before and dates)

   Also relevant to the NAC is the device's security state. The NAC, via the MDM collects additional data such as:
   - (i) Antivirus product and version
   - (ii) Drivers vendors and versions

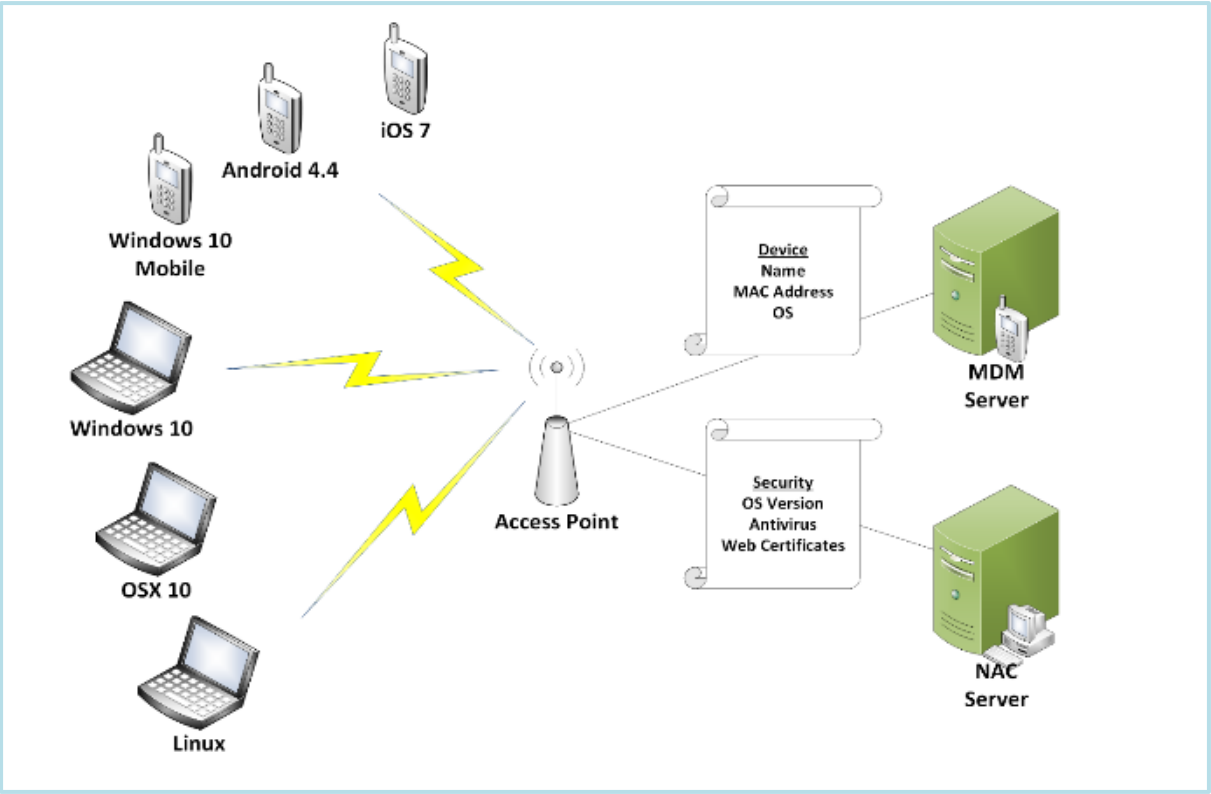

**Figure 8: NAC Mobile Device Management**

On connecting to the WLAN, a mobile device is assigned an IPv4 (or IPv6) address. The device can do nothing else until the NAC assesses and clears it for further authentication based on the data it collects. A device lacking in any way (e.g. outdated virus definitions, discontinued OS versions, compromised device drivers, etc.) is either blocked indefinitely until deficiencies have been met.

Some sophisticated MDMs allow users to register two or more devices to their account, and remove them as necessary.

## 4. Certificates

After the MDM phase, devices can be issued certificates, special customised XML files that are saved in a special directory on the device. Certificates can be generic, or generated using unique device data such as the MAC address or IMEI. This ensures that they cannot be duplicated or copied onto another device which has been blocked or unauthorised. Certificates can also expire and have to be re-issued after a time.

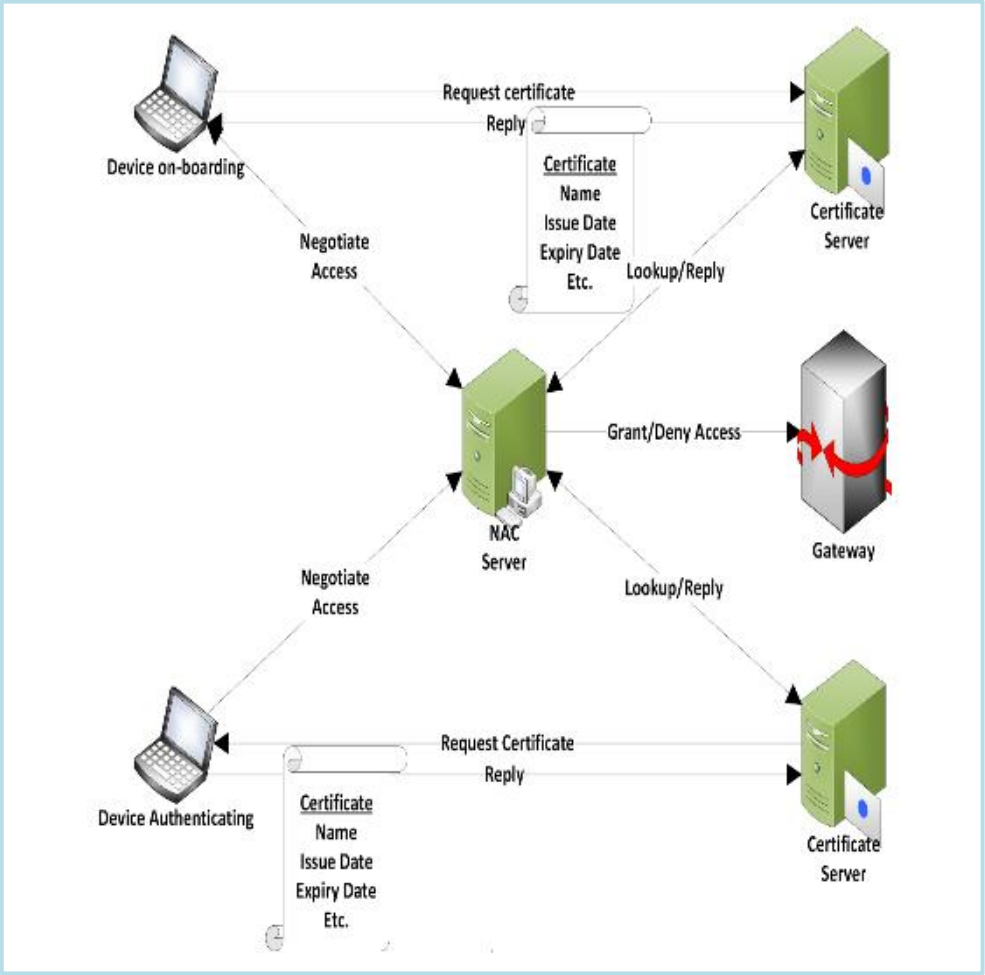

**Figure 9: NAC and Network Certificates**

Figure **9** shows the authentication by certificates, first by an on-boarding device (first joining the network) which is issued a certificate for the first time; and another device whose certificate is being verified before its assigned IP is allowed access through the gateway.

An example of authorizing devices using certificates is implemented by the Cisco Network Setup Assistant [37]. This solution works hand in hand with other Cisco ACMs such as the Cisco ISE. This is how it is implemented in KNUST:

1. The student, having connected to the WLAN, if the device is not already registered, is redirected to a Guest Portal website for registration with their student ID, password and reference number.
2. If the credentials check out, the user is prompted to download a tiny app directly from the school's servers. (There are apps for every platform: iOS, Android, Windows Phone. Desktop OS like Windows, OSX, and Linux may need just applets.)
3. The student runs the app which acquires the device's name, model, IMEI and MAC address (the MAC address is becomes the device ID).
4. The app uses the data to generate and install certificates on the device. Certificates are unique to device and cannot be used on other devices.
5. Each time the student wishes to use a resource on the network, the certificate is required.

## 5. Captive Portal

After device authorization comes student authentication. This is usually the last phase of the authentication process for users attempting to browse the Internet. Because it usually is a website, the captive portal must be accessed with a web browser, an application capable of processing web pages. Otherwise, users will be frustrated in attempting to connect to the Internet yet unable to do so.

Many KNUST students are familiar with the Vodafone captive portal used in Vodafone Internet cafés on campus as well as across the country, and that of the rogue network Cloud Ghana, which is very popular among KNUST students.

The captive portal must be presented to the user the first time the browser attempts to access a website. Since the gateway must present the captive portal to the user, the website is either stored in the gateway itself, or the IP and domain name of the server hosting the captive portal is whitelisted in the gateway's Access Control list. Redirection is achieved using three main techniques:

1. ICMP redirection, which is less common and can easily be bypassed with common IP address spoofing tricks.
2. DNS redirection, whereby the server intercepts a host's DNS lookup and returns the Internet address of the captive portal website; this is called DNS hijacking [38].

The security process of captive portals is quite straightforward. After the student's browser has been redirected to the Captive portal, they must log in with their school issued student ID, password, (and perhaps their reference number as well). The portal contacts a directory server, or a database containing student data. If the credentials check out, the portal contacts the NAC or gateway with the user's device IP to allow said user Internet access. Disconnecting from the Wi-Fi should automatically log a user out as shown in Figure 10.

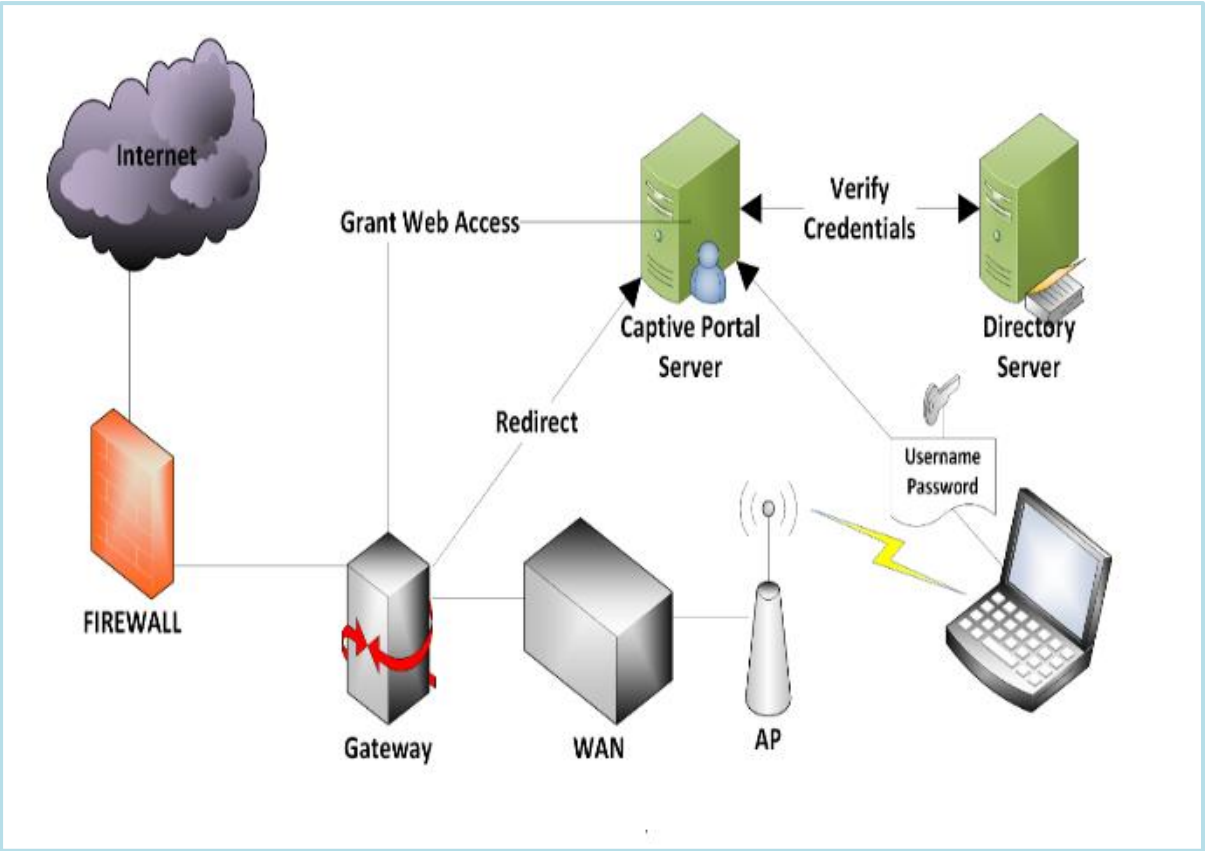

**Figure 10: Captive Portal Authentication Process**

There are so many open source captive portals in existence, such as ChilliSpot, Wifidog, PacketFence and HotSpotPA. All these applications must be installed on a server and properly configured according to the needs and requirements of the target Wi-Fi network.

# 5. SUMMARY

So far it has been proposed that a robust, high-performance WAN-Wi-Fi infrastructure is the foundation of a successful BYOD program. Without a reliable physical network, it is difficult to ensure clients the speed, mobility, security, and QoS they expect to enjoy from a university network; if students reject a poor WLAN will render a BYOD program pointless.

Secondly, it is important to ensure control, security, and convenience in good in a BYOD program in good measures. For control, establish good network monitoring and management solutions. For security, implement a multi-faceted authentication process, which intends to leverage every source of information about the client connecting to the network to use its resources, as in Multi-faceted authentication helps the security system made up of the ACMs (NAC, MDM, and Certificate servers) to recognize contexts and patterns of use and misuse, suspicious behavior and common threats. This eases, at the same time improves, the network administration process. For convenience, the processes of control and security should be abstracted from users, and simplified for them, as much as possible to satisfy usability as summarized in Table 2. Tedious security processes will defeat one of the prime purposes of BYOD: convenience.

**Table 2. Facets of Authentication**

| Security Facet | Relevant Data |
|---|---|
| Who is on the network? | User credentials<br>1. Username and password<br>2. Security questions, etc. |
| What are they using? | User devices<br>1. Device name<br>2. MAC and IP addresses<br>3. IMEI<br>4. OS, Antivirus, etc. |
| Where are they? | Where user accesses the network<br>1. Wired (using a switch or port interface)<br>2. Wireless (connected to an access point) |
| When? | When are they accessing the network, for how long, and how many times<br>1. History logs of device connecting and disconnecting<br>2. Logs of user logins and logouts |
| What can they do? | 1. Access Levels<br>2. Usage (Packet traffic, suspicious behaviour, etc.) |

# 6. CONCLUSION

For many Institutions and Organizations that benefit from computer networking, Wi-Fi is an inevitable choice, and BYOD is the next obvious initiative. While students and employees demand mobility, stakeholders are concerned about security of corporate data. Nevertheless, BYOD is here to stay, and cybersecurity risks have always been a nuisance to network administrators and users long before the introduction of Wi-Fi – risks that can be curtailed or eliminated with tested and proven solutions.

So, regardless of the costs and risks involved, a university' network with a BYOD program is full of benefits that are worth taking the time and resources to thoroughly assess the KNUST's needs, and to properly design, build, manage, and secure a robust physical network and a reliable BYOD program.